\documentclass[lineno,onecolumn,preprint]{jfm}
\usepackage{graphicx}
\usepackage{newtxtext}
\usepackage{newtxmath}
\usepackage{natbib}
\usepackage{hyperref}
\hypersetup{
    colorlinks = true,
    urlcolor   = blue,
    citecolor  = black,
}
\def\kk{{\bf k}}
\def\pp{{\bf p}}
\def\qq{{\bf q}}
\def\bb{{\mathbf{b}_0}}

\newcommand*{\kmax}{k_\mathrm{max}}

\title{Dynamics of fast magnetosonic wave turbulence}

\author{Nicolás P. Müller\aff{1}
  \corresp{\email{nicolas.muller@lpp.polytechnique.fr}} \and
  Sébastien Galtier\aff{1}}

\affiliation{\aff{1}Laboratoire de Physique des Plasmas (LPP), CNRS, École polytechnique, Institut Polytechnique de Paris, Université Paris-Saclay, Sorbonne Université, Observatoire de Paris, 91120 Palaiseau, France}

\begin{document}
\maketitle

\begin{abstract}
Fast magnetosonic waves are among the fundamental oscillation modes of astrophysical  plasmas. To study their dynamics, we carry out numerical simulations of the wave turbulence kinetic equation, which describes the evolution of the energy spectrum of a set of weakly nonlinear fast magnetosonic waves. This kinetic equation, which involves three-wave interactions, has recently been derived from compressible magnetohydrodynamics in the low-$\beta$ limit \citep{Galtier2023}. It has an exact stationary solution, the Kolmogorov-Zakharov spectrum, corresponding to a direct energy cascade. Here we perform free decay simulations of the kinetic equation for which we propose a Kolmogorov-type phenomenology to explain the temporal decay laws of energy and integral length scale. In the forced simulations, we show that the cascade is in fact composed of a mixture of a forward cascade for counter-propagating waves, and a backward cascade for co-propagating waves, with the former being stronger than the latter. The Kolmogorov-Zakharov energy spectrum in $k^{-3/2}$ is found in the radial direction with an anisotropy due to the amplitude that depends on the angle relative to the strong mean magnetic field. We give the analytical expression of the Kolmogorov-Zakharov constant, which is numerically verified in the high Reynolds number limit. Our study provides a theoretical explanation for certain observations in the solar wind plasma \citep{Zhao2022}, where a regime of weak turbulence has been identified for fast magnetosonic waves, alongside a critical balance regime for strong Alfvén wave turbulence. 
\end{abstract}


\section{Introduction}
\label{sec:introduction}

Magnetohydrodynamic (MHD) turbulence plays a central role in astrophysics, whether in the context of the Sun, the solar wind, the interstellar medium, or particle transport \citep{Galtier2016}. MHD turbulence is primarily regarded as a critical balance regime in which high-amplitude Alfvén waves strongly interact with one another, such that the linear time and the nonlinear time are in equilibrium \citep{Goldreich1995}. This critical balance regime helps explain certain observations in space plasmas \citep{Horbury2008,Zhao2024} as well as certain results from direct numerical simulations \citep{Cho2000,Meyrand2016}. The role of the other two modes -- fast and slow magnetosonic waves -- in MHD turbulence has been less studied, partly because Alfvén waves survive in the compressible case and, in a weakly compressible medium such as the solar wind, they remain the dominant mode for the turbulent dynamics. However, the role of fast magnetosonic waves in collisionless plasmas and in particle acceleration including cosmic rays, is well recognized \citep{Brunetti2007,Yan2008,Pongkitiwanichakul2014,Lemoine2021,Hou2025}. 

A recent study of the solar wind conducted at one astronomical unit reveals that, when the plasma $\beta$, defined as the ratio of plasma pressure to magnetic pressure, is small, then the solar wind turbulence can be composed of Alfvénic fluctuations following the phenomenology of critical balance, and of fast magnetosonic fluctuations exhibiting a $k^{-3/2}$-type energy spectrum \citep{Zhao2022}. This observation can be explained using phenomenological arguments based on wave turbulence \citep{Zakharov1970,Cho2002a}. This fast magnetosonic turbulence is often believed to be isotropic, since the spectrum includes only the radial wave number. However, the analytical theory of wave turbulence shows that the amplitude of the spectrum depends on the polar angle, with the amplitude decreasing as the direction approaches the mean magnetic field along the z-direction \citep{Galtier2023}. 
For a long time, the $k^{-3/2}$ energy spectrum (known as the IK-spectrum) was associated with isotropic Alfvén wave turbulence \citep{Iroshnikov1964,Kraichnan1965}. It was later realized that wave turbulence in incompressible MHD is in fact intrinsically anisotropic with a direct energy cascade strictly perpendicular to the mean magnetic field 
\citep{Galtier2000,GaltierC2006,Meyrand2016}. 
It should be noted that, although the IK spectrum is phenomenological in nature, the $k^{-3/2}$ spectrum derived from the theory of fast magnetosonic wave turbulence is an {\it exact} solution to the MHD equations. 

Wave turbulence offers the potential for a deeper understanding of physical systems composed of a set of random waves of weak amplitude interacting nonlinearly. This is explained first by the ability to analytically derive a set of integro-differential equations for the spectral cumulants -- the so-called wave kinetic equations (WKE) -- which are free from the closure problem typically encountered in vortex turbulence. To achieve this natural asymptotic closure, the wave amplitude is used as a small parameter in, \eg a multiple time scale method \citep{Benney1966,Benney1967b,Galtier2024}. Second, exact stationary solutions can be derived from the WKE: these are the so-called Kolmogorov-Zakharov (KZ) spectra, which describe the flux of conserved densities from sources to sinks \citep{Zakharov1967}. Third, these exact solutions correspond to power-law spectra that can be compared to data. The number of experiments, observations, and diagnostics has increased significantly over the past two decades, and today, thanks to high-performance direct numerical simulations and new experiments, wave turbulence has become a leading field in turbulence, from which new fundamental questions have emerged 
\citep{ClarkdiLeoni2016,During2017,Galtier2017,LeReun2017,Pan2017,Yarom2017,Meyrand2018,Hassaini2019,Monsalve2020,Muller2020,Savaro2020,Ricard2021,David2022,Falcon2022,Galtier2022,Griffin2022,Hrabski2022,Onorato2022,Zhang2022,Dematteis2023,Galtier2023b,Lanchon2023,David2024,Gay2024,Kochurin2024,Labarre2024,Ferraro2025,Labarre2025b,Lanchon2025,Shavit2025,Costa2026,Shaltier2026}.

The article is structured as follows. In Section 2, we review the main properties of fast magnetosonic wave turbulence, focusing on the WKE and we provide the analytical expression for the KZ constant. We also describe the simulation code. In Section 3, we present the numerical results, which are organized primarily into two parts: free decay simulations and forced simulations. In Section 4, we develop a discussion and give a conclusion in the last section.

\section{Model and methods}
\label{sec:model}

\subsection{Fast magnetosonic wave turbulence theory}
In compressible magnetohydrodynamics, the dynamics of the plasma can be described by a set of equations governing the evolution of the velocity field, magnetic field, density and pressure. This system supports various types of waves, including Alfv\'en waves, as well as slow  and fast magnetosonic waves. The interactions between these waves and with vortical structures lead to a complex turbulent state. In the limit of a low plasma $\beta$, 
as sometimes observed in the solar wind \citep{Zhao2022},
slow magnetosonic waves can be neglected, and fast magnetosonic waves have the following simple dispersion relation $\omega_F(k) = b_0 k$, where $b_0$ is the Alfvén speed and $k$ the wave number, while for Alfvén waves we still have $\omega_A(k) = b_0 k_\parallel$ (with $\parallel$ the direction along $b_0$). These dependences imply that, at large wave numbers $k$, the two branches can become well separated, suggesting that their respective dynamics are largely decoupled. Direct numerical simulations of incompressible Hall MHD, in which two wave polarizations (left- and right-handed) coexist, have shown that the dynamics of the two polarizations evolve essentially independently \citep{Meyrand2018}.
These arguments provide a physical basis for our simplified approach. Note that the agreement between the theoretical prediction (see below) and the observations provides a posteriori support for the physical validity of the aforementioned assumptions.

In the weakly non-linear regime, where wave interactions dominate over vortex interactions, and by neglecting interactions between different types of waves, we can derive a WKE that describes the statistical properties of fast magnetosonic turbulence and the mechanism of energy transfer between these waves. This WKE, which involves three-wave interactions, takes the following form \citep{Galtier2023}
\begin{equation}
\partial_t E^s_k = 
\label{eq:wke}
\end{equation}
$$
\frac{\pi \epsilon^{2} K_{\theta,\phi}}{32b_0} \int_{\Delta_\perp} \sum_{s_{p} s_{q}} 
\delta(sk+s_pp+s_qq)\frac{s}{pq} 
\left[ sk^3 E^{s_p}_p E^{s_q}_q + s_p p^3 E^s_k E^{s_q}_q + s_q q^3 E^s_k E^{s_p}_p \right]  \mathrm{d} p \mathrm{d} q \, ,  
$$
where $(s,s_p,s_q)=\pm$ denote the directional polarizations associated respectively with the wave vectors $\kk$, $\pp$ and $\qq$ (with $k=\vert \kk \vert$ and the same definition for $p$ and $q$),
$\epsilon$ is a dimensionless nonlinearity parameter (a measure of the wave amplitude, with $0< \epsilon \ll 1$), and $E^s_k=E^s(k)$ is the polarized (kinetic plus magnetic) energy spectrum. 
By definition 
\begin{equation}
K_{\theta,\phi} = \left(\frac{1 + 2 \sin^2 \theta_k}{\sin \theta_k}\right)^2 f_0(\theta_k,\phi_k) \, ,
\end{equation}
is a geometric factor, with $\theta_k$ being the polar angle between the wave vector $\kk$ and the mean magnetic field $\bb=b_0 \mathbf{e}_z$ (with $\vert \mathbf{e}_z \vert = 1$), and $f_0$ is an arbitrary function that depends on the initial condition (with $\phi_k$ being the azimuthal angle). 
The presence of this geometric factor outside the integral reflects a specific property of this semi-dispersive wave turbulence, namely that the cascade develops along rays in three-dimensional space. 
%
This property can be understood using the resonance condition
\begin{gather}
sk + s_pp + s_qq = 0, \\
\kk + \pp + \qq = 0 ,
\label{eq:resonance}
\end{gather}
whose solutions correspond to collinear wave vectors. 
In this case, the integration domain $\Delta_\perp$ reduces to the boundaries of an infinitely long band. 

Note that the derivation of the WKE (\ref{eq:wke}) assumes the existence of weak nonlinearities, but does not necessarily require the random-phase approximation, except at the initial time, as demonstrated rigorously by \cite{Deng2022,Deng2023} and also discussed by \cite{Newell2011} and \cite{Galtier2023b}. Nevertheless, the random-phase approximation is often introduced to simplify the derivation (see, e.g., \cite{ZLF92}). Without this assumption, the derivation becomes considerably more involved, as it may require estimating a large number of oscillatory integrals.

One of the main results of the wave turbulence theory is the derivation of the KZ spectrum, which is an exact stationary solution of equation (\ref{eq:wke}) with a non-zero flux. The total energy spectrum, defined as $E_k=E_k^++E_k^-$, takes the following form \citep{Galtier2023}
\begin{equation}
\label{eq:kz}
E_k = \sqrt{\frac{b_0 \varepsilon}{K_{\theta,\phi}}} C_{KZ} k^{-3/2} \, ,
\end{equation}
where $\varepsilon$ is the positive (and constant) energy flux and $C_{KZ}$ the Kolmogorov-Zakharov constant whose analytical expression that we derived is
\begin{equation}
C_{KZ} = \sqrt{\frac{6}{\pi(\pi-1+4\ln(2))}} \simeq 
0.623 \, .
\end{equation}
It is clear from equation (\ref{eq:kz}) that the amplitude of the spectrum varies as a function of the polar angle $\theta_k$, which gives this spectrum a distinctly anisotropic character never observed elsewhere in anisotropic wave turbulence \citep{Galtier2000,Kuznetsov2001,Galtier2003}. 
The situation is different with regard to the dependence on the azimuthal angle $\phi_k$, which can be neglected when making the statistical assumption of axial symmetry with respect to $\bb$ for the function $f_0$. 
It should be noted that the stronger assumption of isotropy for the function $f_0$ is all the more reasonable when one wishes to study the development of this turbulence without prejudging the form of the resulting anisotropy.

The WKE has a second exact solution, the thermodynamic solution, which corresponds to a zero energy flux. It takes the following form $E_k \sim k^2$. 
The WKE has also a second inviscid invariant, the cross-helicity (the scalar product of velocity and magnetic field), which is by definition \citep{Galtier2023}
\begin{equation}
H_k = (E_k^- - E_k^+) \cos \theta_k . 
\end{equation}
However, the WKE for this quantity has neither a KZ solution nor a thermodynamic solution.

\subsection{Numerical resolution}
In practice, we perform numerical simulations of a modified version of the WKE, namely
\begin{equation}
\partial_t E^s_k = I^s_k + F^s_k - D^s_k ,
\end{equation}
where $I^s_k$ is the right-hand side of Eq.~\eqref{eq:wke}, $D^s_k = \nu k^4 E_k^s$ is a viscous term ($\nu$ is the hyperviscosity) intended to account for small scale dissipation, and $F_k^s$ is a large-scale forcing term defined typically as a Gaussian profile 
$F_k^s = \frac{f^s}{\sqrt{2\pi\Delta_k^2}} \exp[-\frac{(k-k_0)^2}{2\Delta_k^2}]$ 
centered at $k_0$ with a dispersion $\Delta_k$ and an amplitude $f^s$ such that the energy and helicity injection rates are controlled by $f_E = f^+ + f^-$ and $f_H = f^- - f^+$, respectively. 
The inclusion of forcing and dissipation terms allows us to investigate both decaying and forced turbulence scenarios. The implementation of an hyperviscosity instead of a regular viscous term is done to increase the separation between the injection and dissipative scales. 
The collisional integral $I^s_k$ is computed using a linear grid of $N$ points, and the temporal scheme consists of a second-order Runge-Kutta method for the non-linear terms, and an exponential implicit scheme for the viscous term. The numerical implementation is designed to ensure energy conservation in the inviscid limit and to accurately capture the evolution of the energy spectrum and fluxes over time.

\begin{table}
\centering
\begin{tabular}{ c|c c c c c c c c}
   Run & $N$ & $\nu$ & $k_0$ & $E_0$ & $H_0$ & $f_E$ & $f_H$ & $T_f$ \\ 
   \hline
 D1 & $100$  & $10^{-6}$  & $3$ & $1$   & $0$   & $0$   & $0$ & $1.5$ \\  
 D2 & $200$  & $10^{-7}$  & $3$ & $1$   & $0$   & $0$   & $0$ & $1.5$ \\  
 D3 & $400$  & $10^{-8}$  & $3$ & $1$   & $0$   & $0$   & $0$ & $1.5$ \\ 
 D4 & $800$  & $10^{-9}$  & $3$ & $1$   & $0$   & $0$   & $0$ & $1.5$ \\  
 D5 & $1600$ & $10^{-10}$ & $3$ & $1$   & $0$   & $0$   & $0$ & $1.5$ \\  
 D6 & $3200$ & $10^{-11}$ & $3$ & $1$   & $0$   & $0$   & $0$ & $1.5$ \\  
 D7 & $6400$ & $10^{-12}$ & $3$ & $1$   & $0$   & $0$   & $0$ & $1.5$ \\  
 D8 & $800$  & $0$        & $3$ & $1$   & $0$   & $0$   & $0$ & $20$  \\
 D9 & $1600$ & $10^{-10}$ & $10$& $1$   & $0$   & $0$   & $0$ & $1000$  \\
 \hline 
 F1 & $400$  & $10^{-8}$  & $3$ & $0$   & $0$   & $0.5$ & $0$ & $5$ \\
 F2 & $800$  & $10^{-9}$  & $3$ & $0$   & $0$   & $0.5$ & $0$ & $5$ \\  
 F3 & $1600$ & $10^{-10}$ & $3$ & $0$   & $0$   & $0.5$ & $0$ & $5$ \\  
 F4 & $3200$ & $10^{-11}$ & $3$ & $0$   & $0$   & $0.5$ & $0$ & $5$ \\  
 F5 & $6400$ & $10^{-12}$ & $3$ & $0$   & $0$   & $0.5$ & $0$ & $5$ \\  
 F6 & $12800$& $10^{-13}$ & $3$ & $0$   & $0$   & $0.5$ & $0$ & $5$ \\  
 \hline 
 H1 & $1600$ &$2.10^{-10}$& $3$ & $1$   & $0.5$ & $0$   & $0$ & $2$ \\  
 H2 & $800$  & $0$        & $3$ & $1$   & $0.5$ & $0$   & $0$ & $100$ \\
 H3 & $3200$ &$2.10^{-11}$& $3$ & $0$   & $0$   & $1$   &$0.5$& $10$ \\
\end{tabular}
\caption{Table of simulations. The parameter $N$ is the total number of linear collocation points, $\nu$ the hyperviscosity, $k_0$ the center of the Gaussian function used for the forcing or the initial condition, $E_0 = E^+ + E^-$ and $H_0 = E^- - E^+$ the initial amplitudes of the energy and helicity, respectively, $f_E$ and $f_H$ the amplitudes of the energy and helical forcing, respectively, and $T_f$ the final time of the simulation. Runs correspond to decaying (D), forced (F), and helical (H) simulations.}
\label{tab:simulations}
\end{table}

\section{Results}
\label{sec:results}

\subsection{Decaying balanced turbulence}
\label{sec:decaying}

We performed numerical simulations of the WKE \eqref{eq:wke} to investigate the evolution of the energy spectrum and fluxes for various initial conditions and parameter settings. We first consider a balanced ($E_k^+=E_k^-=E_k/2$) decaying turbulence ($F_k^s = 0$), where we start with an initial Gaussian energy spectrum $E_k = E_0 \exp[-(k-k_0)^2/(2 \Delta_k^2)] / \sqrt{2\pi \Delta_k^2}$ centered at $k_0=3$, with $\Delta_k=1$ and initial amplitude $E_0=1$, and let it evolve without external forcing. Runs D1-D7 are evolved up to a final time $T_f=1.5$, shortly after they reach the maximum of dissipation at $t \approx 1.2$. This approach allows us to analyze the development of turbulence. 

\begin{figure}
\centering
\includegraphics[width=.99\textwidth]{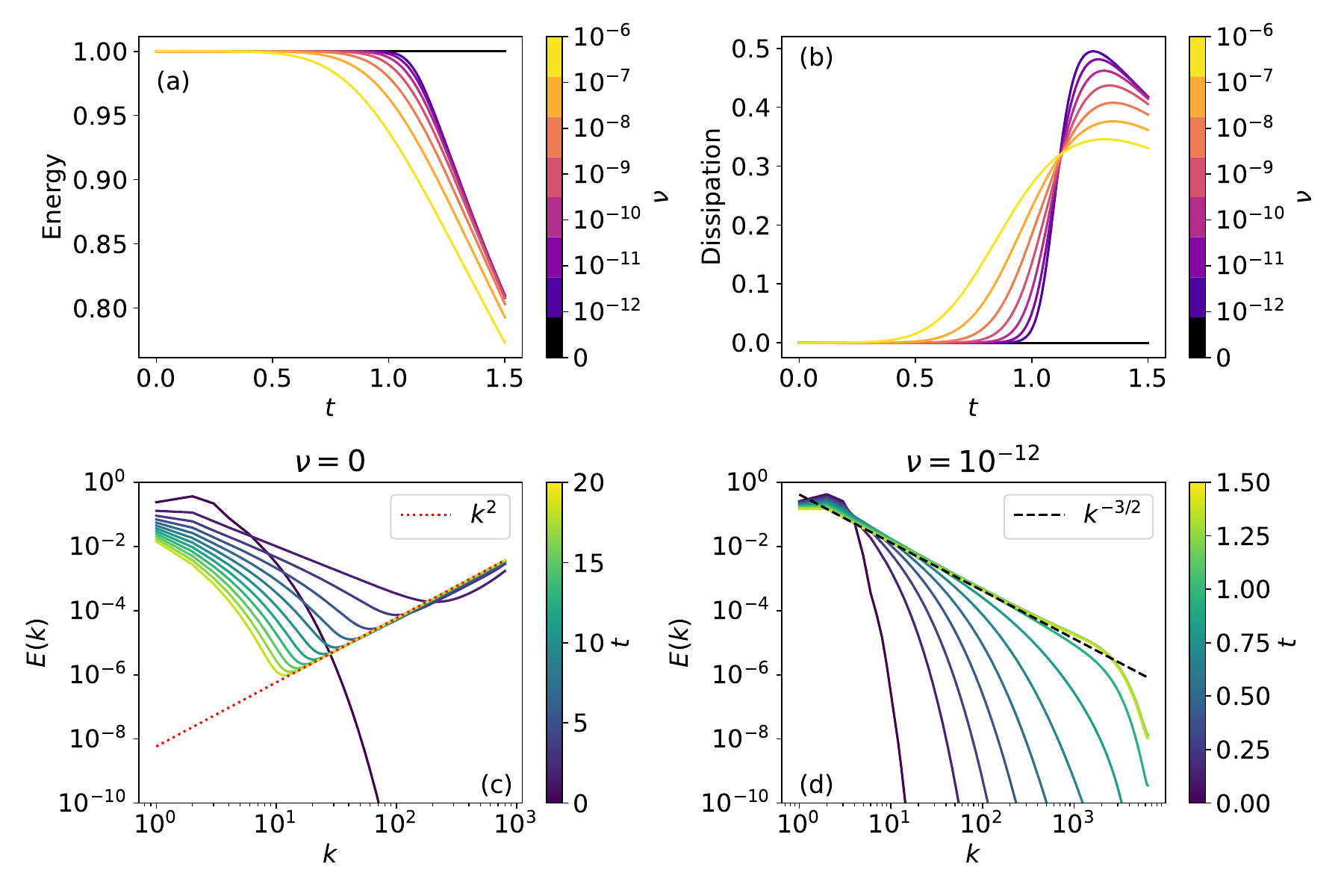}
\caption{Time evolution of the (a) energy and (b) dissipation for different hyperviscosities, from $\nu=10^{-6}$ to $0$ (runs D1-D8 from Table~\ref{tab:simulations}). The temporal evolution of the corresponding energy spectra for (c) $\nu=0$ (run D8) and (d) $\nu = 10^{-12}$ (run D7) are also shown. The scaling $k^2$ (red dotted line) corresponds to the thermodynamic solution, while the scaling $k^{-3/2}$ (black dashed line) corresponds to the KZ solution.}
\label{fig:decaying}
\end{figure}
Figure \ref{fig:decaying} (a) shows the evolution of the total energy $E=\int E_k d \mathrm{k}$ and (b) the dissipation $D=2\int D^+_k d \mathrm{k}$ for different hyperviscosity values, ranging from $\nu=10^{-6}$ to $\nu=0$ (runs D1-D8 from Table~\ref{tab:simulations}). Energy is well conserved in the inviscid case, while in the viscous case we observe a decay of energy and a peak in the dissipation at intermediate times. The energy spectrum at different times is shown in Fig.~\ref{fig:decaying} (c) and (d) for the inviscid and viscous cases, respectively.
In the inviscid case, we observe that the energy spectrum evolves towards a thermodynamic equilibrium \citep{Zakharov1973} at small scales with a $k^2$ scaling, which propagates towards larger scales as the system evolves. In the viscous case, the energy spectrum approaches the KZ solution with a scaling $k^{-3/2}$, indicating a direct cascade of energy to smaller scales. We verified that the thermodynamic solutions have zero energy flux (data not shown), while the KZ solution has a positive flux, as it is discussed in the following. This flux is constant in a very narrow domain, because the system never reaches a stationary state in this decaying simulation. 

The long time evolution and decay of the energy and integral length scale can be predicted using classical Kolmogorov phenomenology \citep{Kolmogorov1941dec,Biskamp1989,Galtier1997,Briard2018}. We will assume that the energy spectrum for $k<k_0$ follows a scaling law \citep{Batchelor1956,Saffman1967}
\begin{equation}
E(k) \sim k^n ,
\end{equation}
and that, in the limit of high Reynolds numbers, the total energy $E = \int E(k) \mathrm{d} k$ and integral scale $\ell_I = \int k^{-1} E(k) \mathrm{d} k / E$ evolve according to a power law with respect to time as 
\begin{gather}
E(t) \sim t^{-\alpha}, \quad \ell_I(t) \sim t^\beta, \label{eq:integral_beta}
\end{gather}
with $\alpha,\beta > 0$.
From these two hypotheses, and assuming that most of the energy is concentrated around the integral length scale, we can derive the following scaling for the energy spectrum
\begin{equation}
E \sim E(k) k \sim k^{n+1} \sim \ell_I^{-(n+1)} \sim t^{- \beta(n+1)}, 
\label{eq:scaling}
\end{equation}
which, combined with the decay of energy, leads to the first relation 
\begin{equation}
\alpha = \beta(n+1).
\label{eq:decay_relation_1}
\end{equation}
A second relation can be derived assuming that the energy decays according to  
\begin{equation}
\frac{\mathrm{d} E}{\mathrm{d} t} \sim t^{-\alpha-1} \sim \frac{E}{\tau_{tr}} ,
\end{equation}
with the transfer time for three-wave interactions $\tau_{tr} = t_{NL}^2 / t_{F} = (\ell / u_\ell)^2 \omega_F \sim b_0 \ell / E$ \citep{Galtier2023}. Note that this is where physically we make a difference with strong (eddy) turbulence for which we have $\tau_{tr} = t_{NL}$. 
This leads to the second relation
\begin{equation}
\alpha + \beta = 1.
\label{eq:decay_relation_2}
\end{equation}
Combining Eqs.~\eqref{eq:decay_relation_1} and \eqref{eq:decay_relation_2}, one obtains that the energy and the integral scale evolve as 
\begin{gather}
  \alpha = \frac{n+1}{n+2} \quad \text{and} \quad  \beta = \frac{1}{n+2} ,
\end{gather}
with $n$ a parameter that depends on the initial condition. According to the phenomenology of decaying turbulence, $n$ can be related to the dimensionality of the system $n = D+1$, which in our case leads to $n=4$ \citep{Batchelor1956}. 
Therefore, our phenomenological predictions for the decay laws are
\begin{equation}
E(t) \sim t^{-5/6} \quad \text{and} \quad
\ell_I(t) \sim t^{1/6} .
\end{equation}

\begin{figure}
\centering
\includegraphics[width=.99\textwidth]{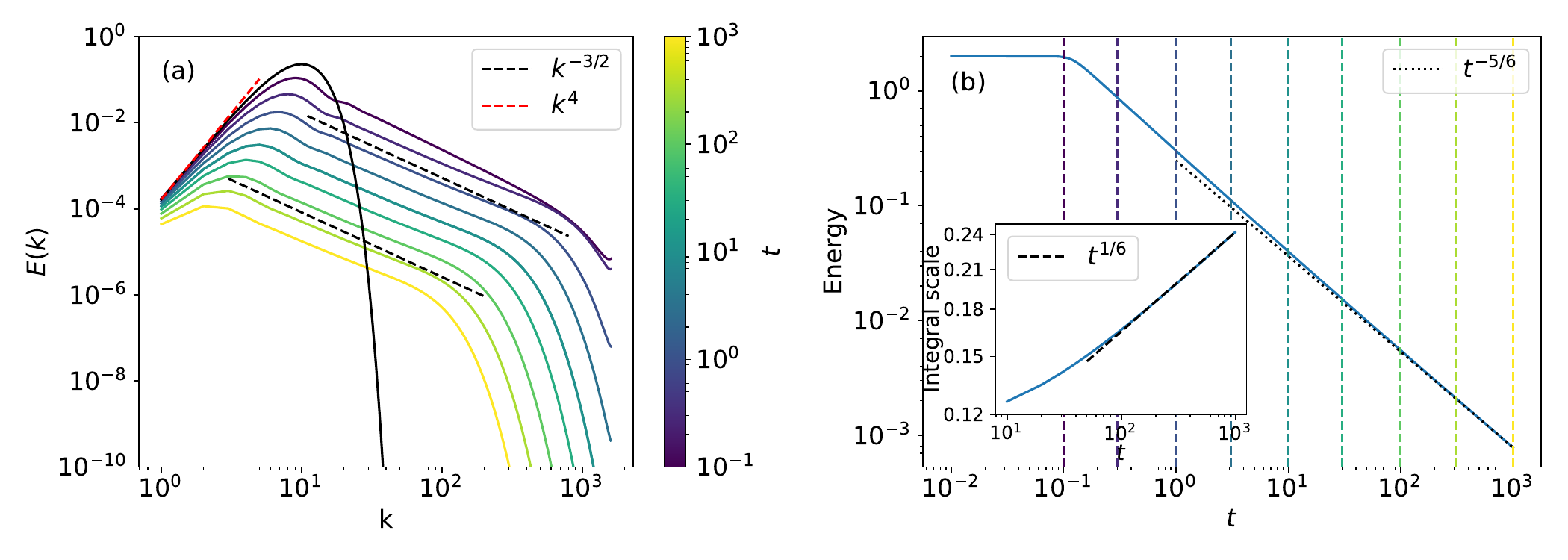}
\caption{Time evolution of the (a) energy spectrum and (b) total energy in logarithmic coordinates (run D9). The initial condition corresponds to $k_0=10$ and $E(k) \sim k^4$ for $k<k_0$. Vertical dashed lines in (b) indicate the times for spectra in (a), and the inset shows the time evolution of the integral scale $\ell_I$. The phenomenological predictions $E \sim t^{-5/6}$ and $\ell_I \sim t^{1/6}$ are plotted in dotted and dashed lines, respectively.}
\label{fig:long_decay}
\end{figure}

Figure \ref{fig:long_decay} (a) shows the long temporal evolution of the energy spectrum for run D9. In this case, the initial condition is $E(k) = \sqrt{512/9\pi} E_0 k_0^{-5} k^4 \exp(-2k^2/k_0^2)$ so that $E(k)\sim k^4$ for $k<k_0$. A self-similar decay of the inertial range is observed, with a spectrum that differs slightly from the $k^{-3/2}$ prediction for wave turbulence. This difference can be attributed to the non-stationary state and the relatively narrow inertial range. 
This point is revisited in Sec.~\ref{sec:discussion}, where we discuss the deviation from the $k^{-3/2}$ scaling in more detail. It is also worth noting that the $k^4$ spectrum at low wave numbers is conserved only over short times. The growth of the integral scale over longer times prevents this conservation.
The temporal decay of energy in Fig.~\ref{fig:long_decay} (b) clearly follows a power-law decay that closely matches the phenomenological prediction. At the same time, the integral length scale (inset) increases with time according to a power law relationship that also closely matches our prediction. 

It is interesting to note that, for weak Alfvén wave turbulence, the phenomenological prediction yields the same estimate for the integral scale, but for energy, it predicts a slower decay in $t^{-2/3}$ due to the strong anisotropy \citep{Bigot2008}. Therefore, fast modes could dissipate faster than Alfvén modes in the weak regime. We arrive at the opposite conclusion for strong incompressible MHD turbulence, for which the decay law follows $t^{-1}$ \citep{Biskamp1999}. 

\subsection{Stationary forced turbulence}
\label{sec:stationary}

\begin{figure}
\centering
\includegraphics[width=.99\textwidth]{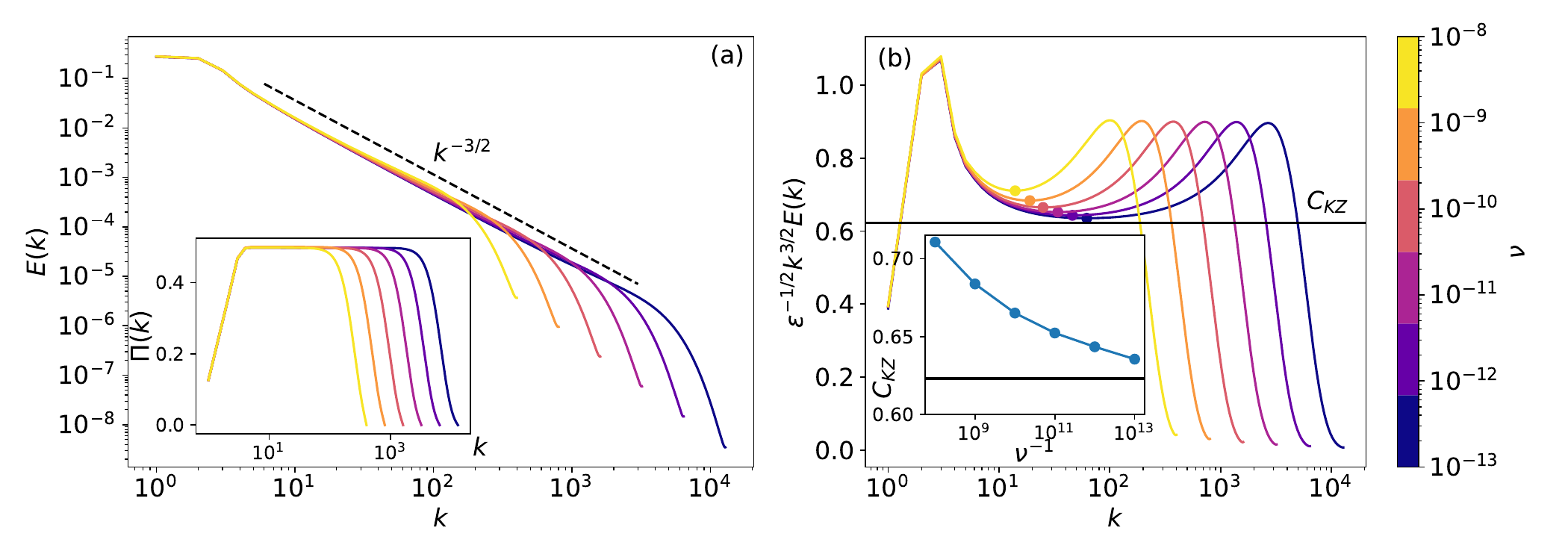}
\caption{(a) Energy spectra, and (b) compensated spectra with the Kolmogorov-Zakharov spectrum \eqref{eq:kz} for different hyperviscosities (or resolutions) corresponding to runs F1-F6. The $k^{-3/2}$ power law is shown in (a) for reference, and the inset shows the corresponding energy fluxes. The inset in (b) shows the minimum values within the inertial range of the compensated spectra, and how these asymptotically approach the theoretical value of the Kolmogorov-Zakharov constant $C_{KZ} \simeq 0.623$.}
\label{fig:stationary}
\end{figure}
A second set of runs are performed with a forcing term in order to reach a stationary state (see runs F1-F6 in Table~\ref{tab:simulations}). We continue to consider a balanced turbulence ($E^+=E^-$). Figure \ref{fig:stationary} (a) shows the energy spectrum for different hyperviscosities. The smaller the hyperviscosity, the more clearly the KZ spectrum in $k^{-3/2}$ is observable. It should be noted that these spectra are plotted for a given angle $\theta_k=\pi/4$. The inset shows the corresponding energy fluxes, defined from the non-linear term as $\Pi_k = - \int_0^k I_{k'} \mathrm{d}k'$. As expected, the KZ spectrum is observed for a constant and positive energy flux $\varepsilon$, confirming the existence of a direct cascade. We note that the value of $\varepsilon$ is consistent with the injection rate of energy controlled by $f_E$. The final property of the KZ spectrum is verified in the right panel where the compensated spectra are shown. It can be seen that the KZ constant is asymptotically approached as the hyperviscosity tends to zero. 

\begin{figure}
\centering
\includegraphics[width=.99\textwidth]{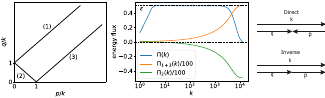}
\caption{Left: integration domain of the wave kinetic equation (\ref{eq:wke}) limited to the three lines (1), (2) and (3). Middle: energy fluxes of run F6 corresponding to lines (1) and (3) (in orange) and to line (2) (in green), normalized by a factor $100$ for visualization. The total energy flux is shown in blue. Right: lines (1) and (3) correspond to two fast waves propagating in opposite directions (top), while line (2) corresponds to two fast waves propagating in the same direction (bottom).}
\label{fig:negative_flux}
\end{figure}
We can go beyond the predictions of the wave turbulence theory by plotting the various contributions to the energy flux. Figure \ref{fig:negative_flux} (left) shows the integration domain for triadic interactions. Usually, the integration is performed along the strip bounded by the three lines (1), (2), and (3), but in the case of semi-dispersive waves, such as fast magnetosonic waves, the integration is restricted to the three lines. On the right, lines (1) and (3) correspond to the interaction (or collision) between two waves propagating in opposite directions, while line (2) corresponds to two waves propagating in the same direction. The middle plot reveals that the energy flux is negative in the latter case and appears to be symmetric to the positive flux found for the first case. Note that fluxes are normalized for visualization. In reality, there is a slight difference, as can be seen in the total energy flux (in blue), which is positive and constant. Therefore, the dominant effect is due to collisions between fast magnetosonic wave packets propagating in opposite directions. 

\subsection{Imbalanced turbulence}
\label{subsec:imbalance}

It is also interesting to examine the situation of imbalanced turbulence where $E^+ \neq E^-$. In this case, energy and cross-helicity are two inviscid invariants of fast magnetosonic wave turbulence, however, unlike energy, the cross-helicity does not have a KZ solution \citep{Galtier2023}. 

\begin{figure}
\centering
\includegraphics[width=.99\textwidth]{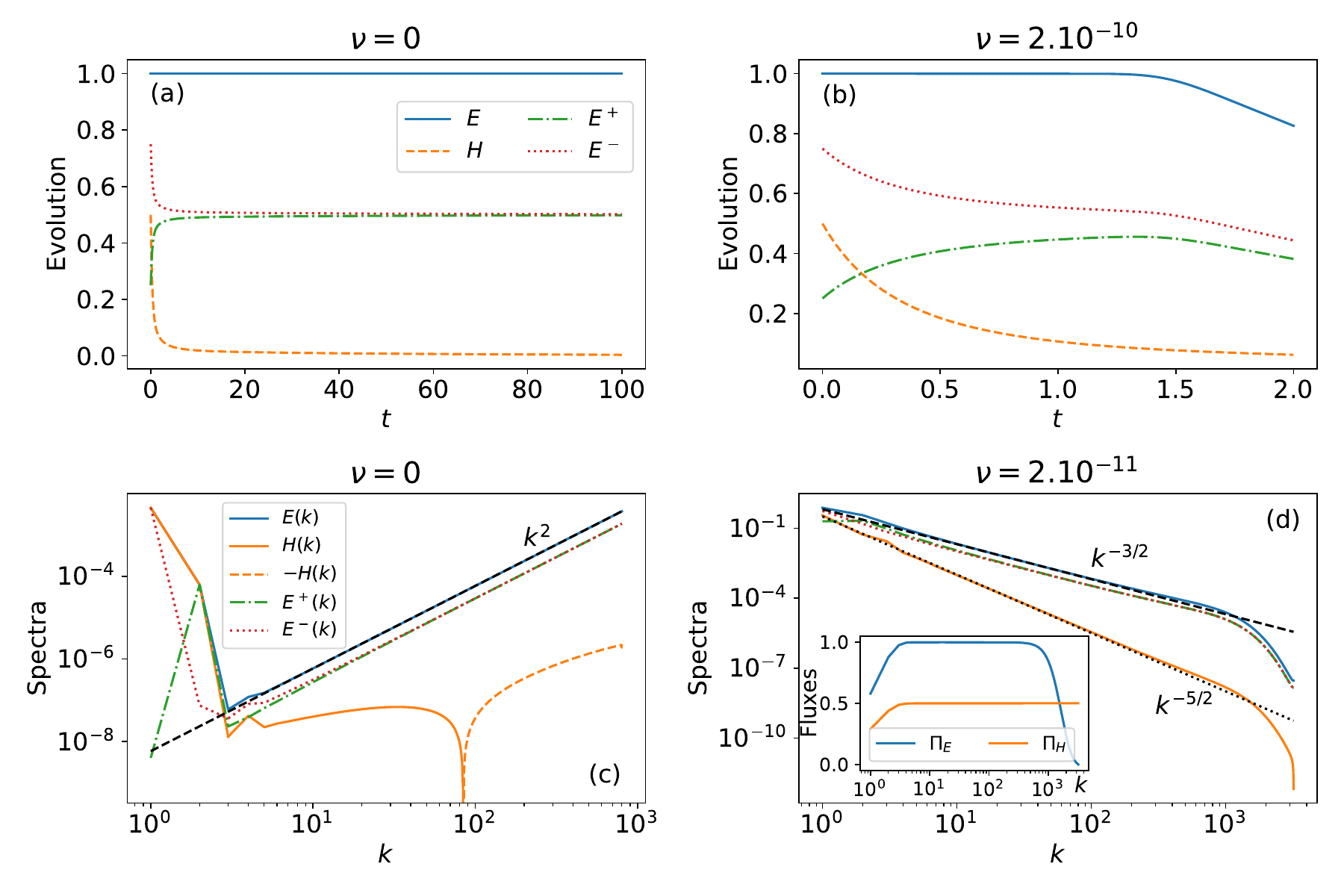}
\caption{Helical simulations. Temporal evolution of the polarized energies $E^+$ (green dash-dotted line), $E^-$ (red dotted line), total energy (blue solid line) and cross-helicity (orange dashed line) for (a) $\nu=0$ (run H2) and (b) $\nu=2 \times 10^{-10}$ (run H1). Spectra at the final time for the inviscid case (c), with the $k^2$ thermodynamic solution (black dashed line) for reference, and (d) for the stationary case (run H3). The inset shows the corresponding energy and helicity fluxes. The spectra are plotted for a given angle $\theta_k=\pi/4$.}
\label{fig:helical}
\end{figure}

To study this situation in more detail, a third set of simulations was performed with $E^- > E^+$ in the decaying case (see runs H1-H2 in Table~\ref{tab:simulations}) and the stationary case (run H3). The results are shown in Figure \ref{fig:helical}. 
Panels (a) and (b) show the temporal variation of the total energy  $E$,  polarized energies $E^\pm$ and cross-helicity $H$. As before, these quantities are plotted along rays for a given polar angle $\theta_k$. 
As expected, during the initial (inviscid) phase of direct cascade, only the total energy is well conserved, not the polarized energies, which tend naturally to converge to the same value. The cross-helicity is also not conserved along rays (i.e. for a given $\theta_k$), which might be surprising, but in fact, only the cross-helicity integrated over $\theta_k$ is conserved. We observe that the system naturally tends to reduce the cross-helicity to zero. This behavior differs from what is observed in weak wave turbulence in incompressible MHD, where the relative helicity tends to increase with time \citep{Galtier2000}. 
Figure \ref{fig:helical}(c) shows the corresponding spectra in the inviscid case, with the thermodynamic solution $k^2$ for reference. As might be expected, only the energy spectra closely follow the analytical solution, while the cross-helicity (its modulus) exhibits a hole due to a change in sign.
In Fig.~\ref{fig:helical}(d), we show the same spectra in the forced case with the KZ solution $k^{-3/2}$ for reference. Clearly, the energy spectra $E^\pm$ follow well the analytical prediction (as expected from the analysis of the WKE), while the cross-helicity spectrum is close to $k^{-5/2}$. This scaling does not correspond to the KZ spectrum and cannot be predicted analytically.
It is interesting to note that in strong Alfvénic turbulence, a similar behavior is found for the correlation energy, which is also defined as the difference between the two polarized energies \citep{Grappin1982,Chandran2008}. But the corresponding spectrum is in $k^{-2}$, while the energy spectrum follows the IK scaling in $k^{-3/2}$.
The inset shows the energy and helicity fluxes for run H3, whose plateaux reach values consistent with their injection rates of $f_E=1$ and $f_H=0.5$. 
Note that the helicity flux does not tend to zero at $k=k_{\mathrm{max}}=N$, as helicity is not conserved along rays. In fact, the helicity flux is not constant in the dissipative scales, instead it slightly increases ($\sim 0.1\%$) to reach a non-zero value at $\kmax$ given by the difference between the two energy polarizations. This difference remains relatively small because the two polarizations approach each other at the dissipative scales, so that their dissipation rates $D_k^+$ and $D_k^-$ are nearly equal. 

\section{Discussion}
\label{sec:discussion}

\begin{figure}
\centering
\includegraphics[width=.99\textwidth]{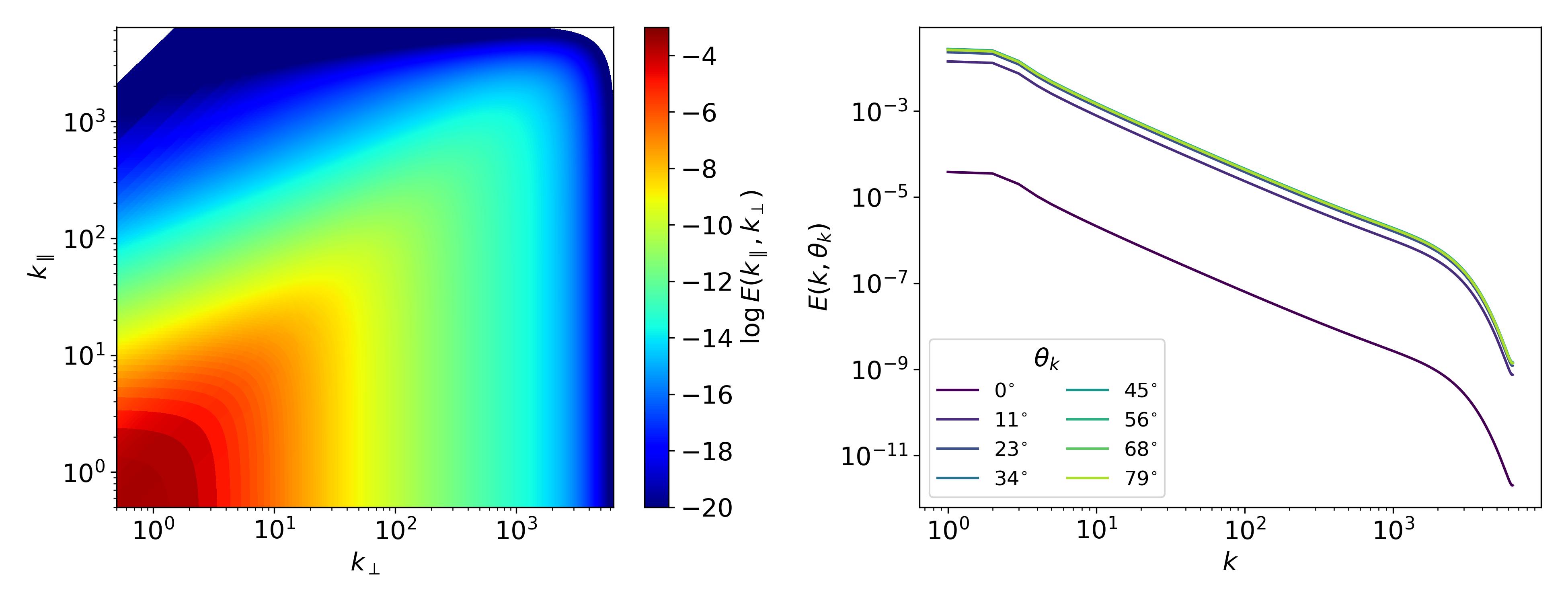}
\caption{Angular dependence of the normalized 2D energy spectrum in a $k_\perp$–$k_\parallel$ plane (left) and for various polar angles $\theta_k$ (right) for run F5. }
\label{fig:angular}
\end{figure}

Even though the KZ energy spectrum follows a $k^{-3/2}$ law, fast magnetosonic wave turbulence is {\it not} isotropic since the amplitude of the total energy spectrum depends on the polar angle $\theta_k$. This property is visible in the normalized two-dimensional (2D) energy spectrum, which can be defined as follows
\begin{equation}
E_{2D}(k,\theta_k) = {\frac{\vert \sin \theta_k \vert}{1+2\sin^2 \theta_k}} E(k) .
\end{equation}
In Figure \ref{fig:angular}, we plot this spectrum in a $k_\perp$–$k_\parallel$ plane (left) and for different polar angles (right) for run F5. As might be expected for fast magnetosonic waves, the smaller the polar angle, the lower the energy. This situation differs significantly from acoustic turbulence, which is fully isotropic. 
It is interesting to note that this result is comparable to a recent observation made in space plasmas \citep{Zhao2024b}, where an improved MHD mode decomposition technique was used to identify the MHD modes. It is thanks to this new technique that the persistence of a weak turbulence regime (at $\beta \sim 1$) has been demonstrated for fast magnetosonic waves, while slow and Alfvén modes transition from a weak turbulence regime to a strong turbulence regime \citep{Zhao2026}. 
As explained by \cite{Galtier2023}, it is also expected, within the range of small $\beta$ values, that strong (critical balance) turbulence for Alfvén modes and weak turbulence for fast modes can coexist. 

\begin{figure}
\centering
\includegraphics[width=.99\textwidth]{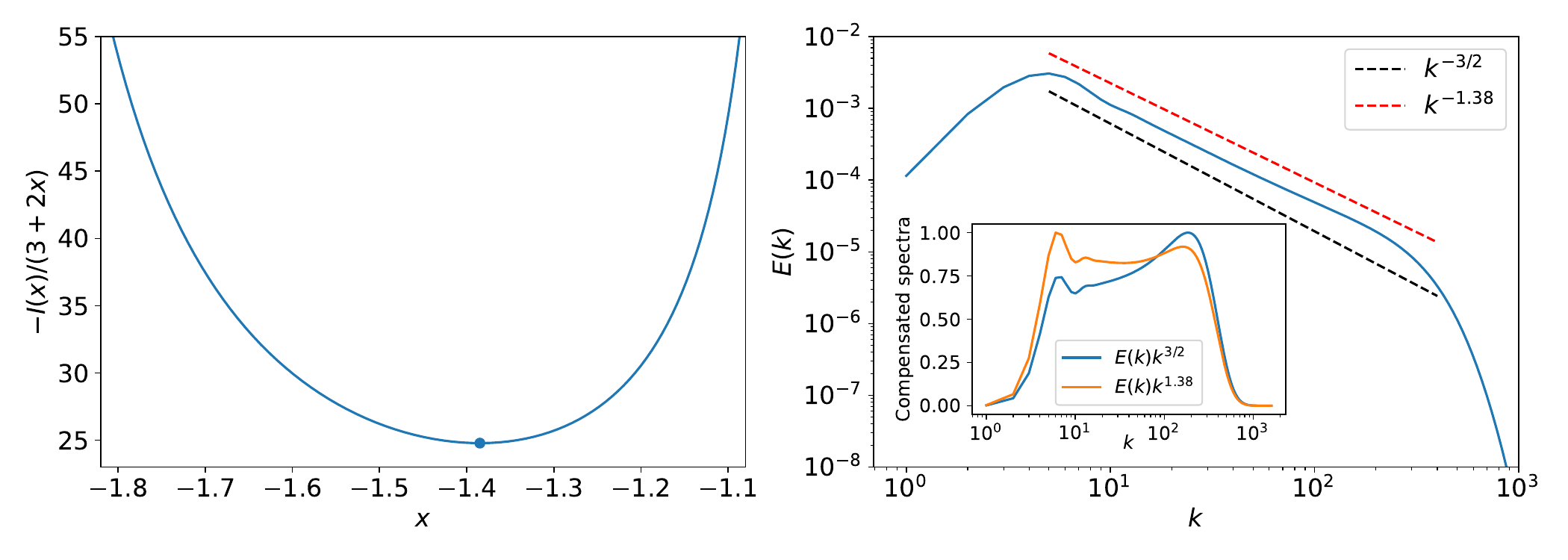}
\caption{(Left) Normalized integral $I(x)$ proportional of the energy flux as in Eq.~\eqref{eq:flux}. (Right) Energy spectrum at $t=10$ for run D9. The KZ (black) and minimal-flux (red) scalings are shown as reference. The inset displays the compensated energy spectra. }
\label{fig:integral}
\end{figure}

Finally, we address the difference observed in the scaling of the energy spectrum with respect to the KZ solution in the long-time decaying regime shown in Fig.~\ref{fig:long_decay}. 
The KZ energy spectrum is obtained by looking for stationary solutions of the form $E_k = C_E k^x$ for the WKE \eqref{eq:wke}. Following the procedure introduced by \cite{Galtier2023}, an expression for the energy flux (Eq.~(5.34) therein) is obtained
\begin{equation}
    \Pi_k = - \frac{\pi K_{\theta, \phi} C_E^2 k^{3+2x}}{32b_0} \frac{I(x)}{3+2x},
    \label{eq:flux}
\end{equation}
where $I(x)$ is an integral resulting from substituting the solution $E_k$ into Eq.~\eqref{eq:wke}. A constant flux is recovered for the value $x=-3/2$, corresponding to the KZ solution. Here, we focus on the non-stationary case, where the flux depends on the wave number $k$. The amplitude of the flux is proportional to the integral $G(x) = -I(x)/(3+2x)$, shown in Fig.~\ref{fig:integral}. For the specific value $x^* = -1.38479$ this integral takes a minimal value, different from the KZ solution.
This is a non-stationary solution of the energy spectrum associated with minimal flux.
Figure~\ref{fig:integral}(b) shows the energy spectra at a late stage of the evolution in the decaying run D9. 
As the system evolves, it seems to relax towards a state that approaches the minimal-flux solution, instead of the KZ one. 
From a physical perspective, this behavior is consistent with the fact that, in the absence of forcing and dissipation, the system relaxes toward a thermodynamic state characterized by zero flux, which is an attractive solution.
The limited resolution of our simulations does not allow us to conclusively determine whether this behavior persists in the system or whether this discrepancy is simply the result of effects associated with a finite Reynolds number. 
This minimal-flux solution suggests the existence of an alternative, non-stationary mechanism for energy redistribution in wave turbulence, distinct from the classical KZ cascade. If confirmed, this could have significant implications for our understanding of energy transfer in weakly nonlinear systems.

\section{Conclusions}
\label{sec:conclusions}

In this study, we show that the properties derived analytically \citep{Galtier2023} are well reproduced by the numerical simulations of the WKE with, in particular, a direct energy cascade characterized by a power law spectrum in $k^{-3/2}$. For the stationary state, we propose an analytical expression of the KZ constant that is asymptotically reached in the limit of large Reynolds numbers. Our numerical study also reveals that the energy cascade is characterized by two opposing fluxes, with a positive flux for interactions between two fast modes propagating in opposite directions and a negative flux for two fast modes propagating in the same direction. We  quantify the anisotropy of fast magnetosonic wave turbulence, which arises from the polar angular dependence of the energy spectrum amplitude. 
We also go beyond the analytical predictions of wave turbulence by studying the decay laws. We show that classical phenomenological arguments apply relatively well, with a power law decay for the energy in $t^{-5/6}$, which is slower than some predictions regarding the energy decay in the case of strong incompressible MHD turbulence \citep{Biskamp1999}. 
The properties highlighted here could lead to a better understanding of the solar wind, which is a decaying, imbalanced turbulence in which the critical balance regime for Alfvén modes and the weak regime for fast magnetosonic modes can coexist \citep{Zhao2022,Zhao2026}. 

It seems important, for the future, to study this regime using direct numerical simulations of compressible MHD at low $\beta$. There are two reasons for this. 
First, with the full equations, all three modes are present and can interact, leading to additional effects not accounted for in the limit considered here, where the fast modes are assumed to evolve independently of the slow modes and Alfvén modes. 
Note, however, that the persistence of a turbulent cascade in the presence of shocks has been confirmed recently with high-resolution plasma kinetic simulations \citep{Hou2025},
as well as signatures of weak turbulence for fast modes \citep{Andres2017,Ugarov2026}. 
Second, in the low $\beta$ limit, the fast modes are semi-dispersive, which a priori constitutes a limiting case for the application of wave turbulence \citep{Benney1966}. However, the same situation is observed for acoustic waves, and, as recently shown by \cite{Kochurin2024} using 3D direct numerical simulations, the wave turbulence regime can be demonstrated. The physical reason advanced is that there are enough collisions between wave packets in 3D to randomize the phases of these wave packets, and allow the wave turbulence regime to emerge \citep{Newell1971,Galtier2023}.  




\backsection[Funding]{NPM and SG are supported by the Simons Foundation (Grant No. 651461, PPC). This project was provided with computer and storage resources by GENCI at CINES thanks to the grant A0200517369 on the supercomputer Adastra GENOA.}

\backsection[Declaration of interests]{The authors report no conflict of interest.}

\backsection[Data availability statement]{
The code and data associated with this work are available from the corresponding author upon request.
}

\backsection[Author ORCIDs]{N. P. M\"uller, https://orcid.org/0000-0003-2204-7873; 
S. Galtier, https://orcid.org/0000-0001-8685-9497}






\bibliographystyle{jfm}
\bibliography{bibliography}

@ARTICLE{Grappin1982,
author = {{Grappin}, R. and {Frisch}, U. and {Pouquet}, A. and {Leorat}, J.},
title = "{Alfvenic fluctuations as asymptotic states of MHD turbulence}",
journal = {Astron. Astrophys.},
year = 1982,
volume = {105},
number = {1},
pages = {6-14}
}

@ARTICLE{Deng2022,
author = {{Deng}, Yu and {Hani}, Zaher},
title = "{Rigorous justification of the wave kinetic theory}",
journal = {arXiv e-prints},
year = 2022,
month = jul,
pages = {arXiv:2207.08358},
doi = {10.48550/arXiv.2207.08358}
}

@ARTICLE{Deng2023,
author = {{Deng}, Yu and {Hani}, Zaher},
title = "{Full derivation of the wave kinetic equation}",
journal = {Inventiones Mathematicae},
year = 2023,
volume = {233},
number = {2},
pages = {543-724},
doi = {10.1007/s00222-023-01189-2}
}

@BOOK{ZLF92,
author = {{Zakharov}, V.E. and {L'Vov}, V.S. and {Falkovich}, G.},
title = {Kolmogorov spectra of turbulence I: Wave turbulence},
publisher = {Springer Series in Nonlinear Dynamics, Berlin: Springer, 1992},
year = 1992}

@ARTICLE{Zhang2022,
author = {{Zhang}, Z. and {Pan}, Y.},
title = "{Numerical investigation of turbulence of surface gravity waves}",
journal = {J. Fluid Mech.},
year = 2022,
volume = {933},
pages = {A58}}

@article{Falcon2022,
author = {Falcon, E. and Mordant, N.},
title = {Experiments in Surface Gravity–Capillary Wave Turbulence},
journal = {Ann. Rev. Fluid Mech.},
volume = {54},
number = {1},
pages = {1-25},
year = {2022}}

@ARTICLE{Savaro2020,
author = {{Savaro}, C. and {Campagne}, A. and {Linares}, M.C. and {Augier}, P. and {Sommeria}, J. and {Valran}, T. and {Viboud}, S. and {Mordant}, N.},
title = "{Generation of weakly nonlinear turbulence of internal gravity waves in the Coriolis facility}",
journal = {Phys. Rev. Fluids},
year = 2020,
volume = {5},
number = {7},
pages = {073801}}

@article{Monsalve2020,
title = {Quantitative Experimental Observation of Weak Inertial-Wave Turbulence},
author = {Monsalve, E. and Brunet, M. and Gallet, B. and Cortet, P.-P.},
journal = {Phys. Rev. Lett.},
volume = {125},
issue = {25},
pages = {254502},
numpages = {6},
year = {2020}}

@ARTICLE{Gay2024,
author = {{Gay}, B. and {Galtier}, S.},
title = "{Gravitational wave turbulence: A multiple time scale approach for quartic wave interactions}",
journal = {Phys. Rev. D},
year = 2024,
volume = {109},
number = {8},
pages = {083531},
doi = {10.1103/PhysRevD.109.083531}
}

@ARTICLE{Lanchon2025,
author = {{Lanchon}, N. and {Boury}, S. and {Cortet}, P.-P.},
title = "{Laboratory observation of internal gravity wave turbulence in a three-dimensional large-scale facility}",
journal = {Phys. Rev. Fluids},
year = 2025,
volume = {10},
number = {8},
pages = {084804},
doi = {10.1103/71dk-9p1c}
}

@ARTICLE{Labarre2025b,
author = {{Labarre}, V. and {Krstulovic}, G. and {Nazarenko}, S.},
title = "{Wave-Kinetic Dynamics of Forced-Dissipated Turbulent Internal Gravity Waves}",
journal = {Phys. Rev. Lett.},
year = 2025,
volume = {135},
number = {1},
pages = {014101},
doi = {10.1103/b82h-z21n}
}

@ARTICLE{Galtier2023b,
author = {{Galtier}, S.},
title = "{A multiple time scale approach for anisotropic inertial wave turbulence}",
journal = {J. Fluid Mech.},
year = 2023,
volume = {974},
pages = {A24},
doi = {10.1017/jfm.2023.825}
}

@ARTICLE{Pan2017,
author = {{Pan}, Y. and {Yue}, D.K.~P.},
title = "{Understanding discrete capillary-wave turbulence using a quasi-resonant kinetic equation}",
journal = {J. Fluid Mech.},
year = 2017,
volume = {816},
pages = {R1},
doi = {10.1017/jfm.2017.106}
}

@ARTICLE{During2017,
author = {{D{\"u}ring}, G. and {Josserand}, C. and {Rica}, S.},
title = "{Wave turbulence theory of elastic plates}",
journal = {Physica D Nonlinear Phenomena},
year = 2017,
volume = {347},
pages = {42-73},
doi = {10.1016/j.physd.2017.01.002}
}

@ARTICLE{David2022,
author = {{David}, V. and {Galtier}, S.},
title = "{Wave turbulence in inertial electron magnetohydrodynamics}",
journal = {J. Plasma Phys.},
year = 2022,
volume = {88},
number = {5},
pages = {905880509},
doi = {10.1017/S0022377822000976}}

@ARTICLE{Lanchon2023,
author = {{Lanchon}, N. and {Mora}, D.O. and {Monsalve}, E. and {Cortet}, P.-P.},
title = "{Internal wave turbulence in a stratified fluid with and without eigenmodes of the experimental domain}",
journal = {Phys. Rev. Fluids},
year = 2023,
volume = {8},
number = {5},
pages = {054802},
doi = {10.1103/PhysRevFluids.8.054802}}

@ARTICLE{Onorato2022,
author = {{Onorato}, M. and {Dematteis}, G. and {Proment}, D. and {Pezzi}, A. and {Ballarin}, M. and {Rondoni}, L.},
title = "{Equilibrium and nonequilibrium description of negative temperature states in a one-dimensional lattice using a wave kinetic approach}",
journal = {Phys. Rev. E},
year = 2022,
volume = {105},
number = {1},
pages = {014206},
doi = {10.1103/PhysRevE.105.014206}}

@ARTICLE{Hrabski2022,
author = {{Hrabski}, A. and {Pan}, Y.},
title = "{On the properties of energy flux in wave turbulence}",
journal = {J. Fluid Mech.},
year = 2022,
volume = {936},
pages = {A47},
doi = {10.1017/jfm.2022.106}}

@ARTICLE{Ferraro2025,
author = {{Ferraro}, M. and {Baudin}, K. and {Gervaziev}, M. and {Fusaro}, A. and {Picozzi}, A. and {Garnier}, J. and {Millot}, G. and {Kharenko}, D. and {Podivilov}, E. and {Babin}, S. and {Mangini}, F. and {Wabnitz}, S.},
title = "{Wave turbulence, thermalization and multimode locking in optical fibers}",
journal = {Physica D Nonlinear Phenomena},
year = 2025,
volume = {481},
pages = {134758},
doi = {10.1016/j.physd.2025.134758}
}

@ARTICLE{LeReun2017,
author = {{Le Reun}, T. and {Favier}, B. and {Barker}, A.J. and {Le Bars}, M.},
title = "{Inertial Wave Turbulence Driven by Elliptical Instability}",
journal = {Phys. Rev. Lett.},
year = 2017,
volume = {119},
number = {3},
pages = {034502}}

@ARTICLE{Galtier2017,
author = {{Galtier}, S. and {Nazarenko}, S.V.},
title = "{Turbulence of Weak Gravitational Waves in the Early Universe}",
journal = {Phys. Rev. Lett.},
year = 2017,
volume = {119},
number = {22},
pages = {221101}}

@ARTICLE{Hassaini2019,
author = {{Hassaini}, R. and {Mordant}, N. and {Miquel}, B. and {Krstulovic}, G. and {D{\"u}ring}, G.},
title = "{Elastic weak turbulence: From the vibrating plate to the drum}",
journal = {Phys. Rev. E},
year = 2019,
volume = {99},
number = {3},
pages = {033002}}

@ARTICLE{Yarom2017,
author = {{Yarom}, E. and {Salhov}, A. and {Sharon}, E.},
title = "{Experimental quantification of nonlinear time scales in inertial wave rotating turbulence}",
journal = {Phys. Rev. Fluids},
year = 2017,
volume = {2},
number = {12},
pages = {122601},
doi = {10.1103/PhysRevFluids.2.122601}
}

@ARTICLE{Dematteis2023,
author = {{Dematteis}, G. and {Lvov}, Y.V.},
title = "{The structure of energy fluxes in wave turbulence}",
journal = {J. Fluid Mech.},
year = 2023,
volume = {954},
pages = {A30},
doi = {10.1017/jfm.2022.995}}

@ARTICLE{Ricard2021,
author = {{Ricard}, G. and {Falcon}, E.},
title = "{Experimental quasi-1D capillary-wave turbulence}",
journal = {EPL (Europhysics Letters)},
year = 2021,
volume = {135},
number = {6},
pages = {64001},
doi = {10.1209/0295-5075/ac2751}}

@ARTICLE{Labarre2024,
author = {{Labarre}, V. and {Lanchon}, N. and {Cortet}, P.-P. and {Krstulovic}, G. and {Nazarenko}, S.},
title = "{On the kinetics of internal gravity waves beyond the hydrostatic regime}",
journal = {J. Fluid Mech.},
year = 2024,
volume = {998},
pages = {A17},
doi = {10.1017/jfm.2024.802}}

@ARTICLE{David2024,
author = {{David}, V. and {Galtier}, S. and {Meyrand}, R.},
title = "{Monofractality in the Solar Wind at Electron Scales: Insights from Kinetic Alfv{\'e}n Waves Turbulence}",
journal = {Phys. Rev. Lett.},
year = 2024,
volume = {132},
number = {8},
pages = {085201},
doi = {10.1103/PhysRevLett.132.085201}}

@ARTICLE{Shavit2025,
author = {{Shavit}, M. and {B{\"u}hler}, O. and {Shatah}, J.},
title = "{Turbulent Spectrum of 2D Internal Gravity Waves}",
journal = {Phys. Rev. Lett.},
year = 2025,
month = feb,
volume = {134},
number = {5},
pages = {054101},
doi = {10.1103/PhysRevLett.134.054101}
}

@article{Andres2017,
  title = {Interplay between {{Alfv\'en}} and Magnetosonic Waves in Compressible Magnetohydrodynamics Turbulence},
  author = {Andr{\'e}s, N. and Clark Di Leoni, P. and Mininni, P. D. and Dmitruk, P. and Sahraoui, F. and Matthaeus, W. H.},
  year = 2017,
  month = oct,
  journal = {Physics of Plasmas},
  volume = {24},
  number = {10},
  pages = {102314},
  issn = {1070-664X, 1089-7674},
  doi = {10.1063/1.4997990},
  urldate = {2025-09-12},
  langid = {english}
}

@ARTICLE{Benney1966,
author = {{Benney}, D.J. and {Saffman}, P.G.},
title = "{Nonlinear Interactions of Random Waves in a Dispersive Medium}",
journal = {Proc. R. Soc. Lond. A},
year = 1966,
volume = {289},
number = {1418},
pages = {301-320}}

@article{Benney1967b,
Author = {Benney, D.J. and Newell, A.C.},
Title = {{Sequential time closures for interacting random waves}},
Journal = {{J. Math. Phys.}},
Year = {{1967}},
Volume = {{46}},
Number = {{4}},
Pages = {{363-392}}}

@ARTICLE{Bigot2008,
author = {{Bigot}, B. and {Galtier}, S. and {Politano}, H.},
title = "{Energy Decay Laws in Strongly Anisotropic Magnetohydrodynamic Turbulence}",
journal = {Phys. Rev. Lett.},
year = 2008,
volume = {100},
number = {7},
pages = {074502},
doi = {10.1103/PhysRevLett.100.074502}
}

@ARTICLE{Zakharov1973,
author = {{Zakharov}, V.~E. and {Musher}, S.~L.},
title = "{Kolmogorov spectrum in a system of nonlinear oscillators}",
journal = {Soviet Physics Doklady},
year = 1973,
volume = {18},
pages = {240}
}

@ARTICLE{Brunetti2007,
author = {{Brunetti}, G. and {Lazarian}, A.},
title = "{Compressible turbulence in galaxy clusters: physics and stochastic particle re-acceleration}",
journal = {MNRAS},
year = 2007,
month = jun,
volume = {378},
number = {1},
pages = {245-275},
doi = {10.1111/j.1365-2966.2007.11771.x}}

@article{Chandran2008,
  title = {Strong {{Anisotropic MHD Turbulence}} with {{Cross Helicity}}},
  author = {Chandran, Benjamin D. G.},
  year = 2008,
  month = sep,
  journal = {The Astrophysical Journal},
  volume = {685},
  number = {1},
  pages = {646--658},
  issn = {0004-637X, 1538-4357},
  doi = {10.1086/589432},
  urldate = {2025-09-24},
  langid = {english}
}

@article{Cho2002a,
  title = {Compressible {{Sub-Alfv\'enic MHD Turbulence}} in {{Low-}} {$\beta$} {{Plasmas}}},
  author = {Cho, J. and Lazarian, A.},
  year = 2002,
  month = may,
  journal = {Phys. Rev. Lett.},
  volume = {88},
  number = {24},
  pages = {245001},
  issn = {0031-9007, 1079-7114},
  doi = {10.1103/PhysRevLett.88.245001},
  urldate = {2025-09-15},
  copyright = {http://link.aps.org/licenses/aps-default-license},
  langid = {english}
}

@ARTICLE{Cho2000,
author = {{Cho}, J. and {Vishniac}, E.T.},
title = "{The Anisotropy of Magnetohydrodynamic Alfv{\'e}nic Turbulence}",
journal = {Astrophys. J.},
year = 2000,
volume = {539},
number = {1},
pages = {273-282},
doi = {10.1086/309213}
}

@ARTICLE{ClarkdiLeoni2016,
author = {{Clark di Leoni}, P. and {Mininni}, P.D.},
title = "{Quantifying resonant and near-resonant interactions in rotating turbulence}",
journal = {J. Fluid Mech.},
year = 2016,
volume = {809},
pages = {821-842},
doi = {10.1017/jfm.2016.713}
}

@ARTICLE{Shaltier2026,
author = {{Shaltiel}, Omri and {Gat}, Omri and {Sharon}, Eran},
title = "{Evidence for Wave Turbulence Spectra in Rotating Turbulence}",
journal = {Phys. Rev. Lett.},
year = 2026,
volume = {137},
number = {8},
pages = {084001},
doi = {10.1103/33dv-bfcx}
}

@ARTICLE{Batchelor1956,
author = {{Batchelor}, G.~K. and {Proudman}, I.},
title = "{The Large-Scale Structure of Homogeneous Turbulence}",
journal = {Phil. Trans. Royal Soc. London Series A},
year = 1956,
volume = {248},
number = {949},
pages = {369-405},
doi = {10.1098/rsta.1956.0002},
}

@ARTICLE{Saffman1967,
author = {{Saffman}, P.~G.},
title = "{Note on Decay of Homogeneous Turbulence}",
journal = {Phys. Fluids},
year = 1967,
volume = {10},
number = {6},
pages = {1349-1349},
doi = {10.1063/1.1762284}
}

@ARTICLE{Biskamp1989,
author = {{Biskamp}, D. and {Welter}, H.},
title = "{Dynamics of decaying two-dimensional magnetohydrodynamic turbulence}",
journal = {Phys. Fluids B},
year = 1989,
volume = {1},
number = {10},
pages = {1964-1979},
doi = {10.1063/1.859060}
}

@ARTICLE{Biskamp1999,
author = {{Biskamp}, D. and {M{\"u}ller}, W.-C.},
title = "{Decay Laws for Three-Dimensional Magnetohydrodynamic Turbulence}",
journal = {Phys. Rev. Lett.},
year = 1999,
volume = {83},
number = {11},
pages = {2195-2198},
doi = {10.1103/PhysRevLett.83.2195}
}

@ARTICLE{Galtier1997,
author = {{Galtier}, S. and {Politano}, H. and {Pouquet}, A.},
title = "{Self-Similar Energy Decay in Magnetohydrodynamic Turbulence}",
journal = {Phys. Rev. Lett.},
year = 1997,
volume = {79},
number = {15},
pages = {2807-2810},
doi = {10.1103/PhysRevLett.79.2807}
}

@ARTICLE{Briard2018,
author = {{Briard}, A. and {Gomez}, T.},
title = "{The decay of isotropic magnetohydrodynamics turbulence and the effects of cross-helicity}",
journal = {J. Plasma Phys.},
year = 2018,
volume = {84},
number = {1},
pages = {905840110},
doi = {10.1017/S0022377818000120}
}

@article{Galtier2000,
  title = {A Weak Turbulence Theory for Incompressible Magnetohydrodynamics},
  author = {Galtier, S. and Nazarenko, S. V. and Newell, A. C. and Pouquet, A.},
  year = 2000,
  month = jun,
  journal = {Journal of Plasma Physics},
  volume = {63},
  number = {5},
  pages = {447--488},
  issn = {0022-3778, 1469-7807},
  doi = {10.1017/S0022377899008284},
  urldate = {2026-01-14},
  copyright = {https://www.cambridge.org/core/terms},
  langid = {english}
}

@ARTICLE{Galtier2003,
author = {{Galtier}, S.},
title = "{Weak inertial-wave turbulence theory}",
journal = {Phys. Rev. E},
year = "2003",
volume = {68},
number = {1},
doi = {10.1103/PhysRevE.68.015301},
pages = {015301}}

@book{Galtier2016,
  title = {Introduction to {{Modern Magnetohydrodynamics}}},
  author = {Galtier, S.},
  year = 2016,
  month = sep,
  edition = {1},
  publisher = {Cambridge University Press},
  doi = {10.1017/CBO9781316665961},
  urldate = {2023-11-14},
  isbn = {978-1-107-15865-8 978-1-316-66596-1},
  langid = {english}
}

@book{Galtier2022,
  title = {Physics of {{Wave Turbulence}}},
  author = {Galtier, S.},
  year = 2022,
  month = dec,
  edition = {1},
  publisher = {Cambridge University Press},
  doi = {10.1017/9781009275880},
  urldate = {2025-09-17},
  copyright = {https://www.cambridge.org/core/terms},
  isbn = {978-1-009-27588-0 978-1-009-27589-7}
}

@article{Galtier2023,
  title = {Fast Magneto-Acoustic Wave Turbulence and the {{Iroshnikov}}--{{Kraichnan}} Spectrum},
  author = {Galtier, S.},
  year = 2023,
  month = apr,
  journal = {Journal of Plasma Physics},
  volume = {89},
  number = {2},
  pages = {905890205},
  issn = {0022-3778, 1469-7807},
  doi = {10.1017/S0022377823000259},
  urldate = {2025-06-04},
  langid = {english}
}

@article{Galtier2024,
  title = {Wave Turbulence: A Solvable Problem Applied to the {{Navier}}--{{Stokes}} Equations},
  shorttitle = {Wave Turbulence},
  author = {Galtier, S.},
  year = 2024,
  journal = {Comptes Rendus. Physique},
  volume = {25},
  number = {G1},
  pages = {433--455},
  doi = {10.5802/crphys.221}
}

@article{Goldreich1995,
  title = {Toward a Theory of Interstellar Turbulence. 2: {{Strong}} Alfvenic Turbulence},
  shorttitle = {Toward a Theory of Interstellar Turbulence. 2},
  author = {Goldreich, P. and Sridhar, S.},
  year = 1995,
  month = jan,
  journal = {Astrophys. J.},
  volume = {438},
  pages = {763},
  issn = {0004-637X, 1538-4357},
  doi = {10.1086/175121},
  urldate = {2025-09-24},
  langid = {english}
}

@article{Griffin2022,
  title = {Energy {{Spectrum}} of {{Two-Dimensional Acoustic Turbulence}}},
  author = {Griffin, Adam and Krstulovic, Giorgio and L'vov, Victor S. and Nazarenko, Sergey},
  year = 2022,
  month = jun,
  journal = {Physical Review Letters},
  volume = {128},
  number = {22},
  pages = {224501},
  issn = {0031-9007, 1079-7114},
  doi = {10.1103/PhysRevLett.128.224501},
  urldate = {2025-11-24},
  langid = {english}
}

@ARTICLE{Horbury2008,
author = {{Horbury}, T.S. and {Forman}, M. and {Oughton}, S.},
title = "{Anisotropic Scaling of Magnetohydrodynamic Turbulence}",
journal = {Phys. Rev. Lett.},
year = 2008,
volume = {101},
number = {17},
pages = {175005},
doi = {10.1103/PhysRevLett.101.175005}}

@article{Hou2025,
  title = {Energy cascade and damping in fast-mode compressible turbulence},
  author = {Hou, C. and Yan, H. and Pavaskar, P.},
  year = 2025,
  journal = {Astrophys. J. Lett.},
  volume = {992},
  pages = {L28}
}

@article{Kochurin2024,
  title = {Three-{{Dimensional Acoustic Turbulence}}: {{Weak Versus Strong}}},
  shorttitle = {Three-{{Dimensional Acoustic Turbulence}}},
  author = {Kochurin, E. A. and Kuznetsov, E. A.},
  year = 2024,
  month = nov,
  journal = {Physical Review Letters},
  volume = {133},
  number = {20},
  pages = {207201},
  issn = {0031-9007, 1079-7114},
  doi = {10.1103/PhysRevLett.133.207201},
  urldate = {2025-06-10},
  langid = {english}
}

@article{Kolmogorov1941dec,
  title = {On degeneration of isotropic turbulence in an incompressible viscous fluid},
  author = {Kolmogorov, A.N.},
  year = 1941,
  journal = {Doklady Akademii Nauk},
  volume = {31},
  pages = {538--540}
}

@article{Kuznetsov2001,
  title = {Weak Magnetohydrodynamic Turbulence of a Magnetized Plasma},
  author = {Kuznetsov, E. A.},
  year = 2001,
  month = nov,
  journal = {Journal of Experimental and Theoretical Physics},
  volume = {93},
  number = {5},
  pages = {1052--1064},
  issn = {1063-7761, 1090-6509},
  doi = {10.1134/1.1427116},
  urldate = {2025-09-24},
  copyright = {http://www.springer.com/tdm},
  langid = {english}
}

@ARTICLE{Lemoine2021,
author = {{Lemoine}, M.},
title = "{Particle acceleration in strong MHD turbulence}",
journal = {Phys. Rev. D},
year = 2021,
month = sep,
volume = {104},
number = {6},
pages = {063020},
doi = {10.1103/PhysRevD.104.063020}}

@article{Meyrand2016,
  title = {Direct {{Evidence}} of the {{Transition}} from {{Weak}} to {{Strong Magnetohydrodynamic Turbulence}}},
  author = {Meyrand, R. and Galtier, S. and Kiyani, K.H.},
  year = 2016,
  month = mar,
  journal = {Physical Review Letters},
  volume = {116},
  number = {10},
  pages = {105002},
  issn = {0031-9007, 1079-7114},
  doi = {10.1103/PhysRevLett.116.105002},
  urldate = {2025-09-15},
  copyright = {http://link.aps.org/licenses/aps-default-license},
  langid = {english}
}

@article{Meyrand2018,
  title = {Coexistence of {{Weak}} and {{Strong Wave Turbulence}} in {{Incompressible Hall Magnetohydrodynamics}}},
  author = {Meyrand, R. and Kiyani, K.H. and G{\"u}rcan, O.D. and Galtier, S.},
  year = 2018,
  month = sep,
  journal = {Physical Review X},
  volume = {8},
  number = {3},
  pages = {031066},
  issn = {2160-3308},
  doi = {10.1103/PhysRevX.8.031066},
  urldate = {2025-10-27},
  langid = {english}
}

@article{Muller2020,
  title = {Kolmogorov and {{Kelvin}} Wave Cascades in a Generalized Model for Quantum Turbulence},
  author = {M{\"u}ller, N.P. and Krstulovic, G.},
  year = 2020,
  month = oct,
  journal = {Physical Review B},
  volume = {102},
  number = {13},
  pages = {134513},
  issn = {2469-9950, 2469-9969},
  doi = {10.1103/PhysRevB.102.134513},
  urldate = {2021-05-10},
  copyright = {All rights reserved},
  langid = {english}
}

@ARTICLE{Newell2011,
author = {{Newell}, Alan C. and {Rumpf}, Benno},
title = "{Wave Turbulence}",
journal = {Annual Review of Fluid Mechanics},
year = 2011,
volume = {43},
number = {1},
pages = {59-78},
doi = {10.1146/annurev-fluid-122109-160807},
}

@ARTICLE{Pongkitiwanichakul2014,
author = {{Pongkitiwanichakul}, P. and {Chandran}, B.D.~G.},
title = "{Stochastic Acceleration of Electrons by Fast Magnetosonic Waves in Solar Flares: The Effects of Anisotropy in Velocity and Wavenumber Space}",
journal = {Astrophys. J.},
year = 2014,
month = nov,
volume = {796},
number = {1},
pages = {45},
doi = {10.1088/0004-637X/796/1/45}
}

@ARTICLE{Newell1971,
author = {{Newell}, A.~C. and {Aucoin}, P.~J.},
title = "{Semidispersive wave systems}",
journal = {J. Fluid Mech.},
year = 1971,
volume = {49},
pages = {593-609},
doi = {10.1017/S0022112071002271}
}

@ARTICLE{Ugarov2026,
author = {{Ugarov}, P. and {Zhdankin}, V. and {Arr{\`o}}, G.},
title = "{Fast Magnetosonic Turbulence in Two-Dimensional Relativistic Plasmas}",
journal = {arXiv e-prints},
year = 2026,
pages = {arXiv:2604.04276},
doi = {10.48550/arXiv.2604.04276},
archivePrefix = {arXiv},
eprint = {2604.04276},
}

@ARTICLE{Yan2008,
author = {{Yan}, H. and {Lazarian}, A. and {Petrosian}, V.},
title = "{Particle Acceleration by Fast Modes in Solar Flares}",
journal = {Astrophys. J.},
year = 2008,
month = sep,
volume = {684},
number = {2},
pages = {1461-1468},
doi = {10.1086/589962}
}

@ARTICLE{Zakharov1967,
author = {{Zakharov}, V.E. and {Filonenko}, N.N.},
title = "{Weak turbulence of capillary waves}",
journal = {J. Applied Mech. Tech. Phys.},
year = "1967",
volume = {8},
number = {5},
doi = {10.1007/BF00915178},
pages = {37-40}}

@article{Zakharov1970,
  title = {Spectrum of {{Acoustic Turbulence}}},
  author = {Zakharov, V. E. and Sagdeev, R. Z.},
  year = 1970,
  month = nov,
  journal = {Soviet Physics Doklady},
  volume = {15},
  pages = {439},
  urldate = {2025-12-30}
}

@article{Zhao2022,
  title = {Multispacecraft {{Analysis}} of the {{Properties}} of {{Magnetohydrodynamic Fluctuations}} in {{Sub-Alfv\'enic Solar Wind Turbulence}} at 1 Au},
  author = {Zhao, S. Q. and Yan, Huirong and Liu, Terry Z. and Liu, Mingzhe and Wang, Huizi},
  year = 2022,
  month = oct,
  journal = {Astrophys. J.},
  volume = {937},
  number = {2},
  pages = {102},
  issn = {0004-637X, 1538-4357},
  doi = {10.3847/1538-4357/ac822e},
  urldate = {2025-09-24},
  langid = {english}
}

@ARTICLE{Zhao2024,
author = {{Zhao}, S. and {Yan}, H. and {Liu}, T.Z. and {Yuen}, K.H. and {Wang}, H.},
title = "{Identification of the weak-to-strong transition in Alfv{\'e}nic turbulence from space plasma}",
journal = {Nature Astronomy},
year = 2024,
month = jun,
volume = {8},
pages = {725-731},
doi = {10.1038/s41550-024-02249-0}
}

@ARTICLE{Zhao2024b,
author = {{Zhao}, S. and {Yan}, H. and {Liu}, T.Z. and {Yuen}, K.H. and {Shi}, M.},
title = "{Small-amplitude Compressible Magnetohydrodynamic Turbulence Modulated by Collisionless Damping in Earth's Magnetosheath: Observation Matches Theory}",
journal = {Astrophys. J.},
year = 2024,
volume = {962},
number = {1},
pages = {89},
doi = {10.3847/1538-4357/ad132e}
}

@ARTICLE{Zhao2026,
author = {{Zhao}, S. and {Yan}, H. and {Liu}, T.Z. and {Hou}, C. and {Yuen}, K.H.},
title = "{Spatiotemporal Properties of Compressible Magnetohydrodynamic Turbulence from Space Plasma}",
journal = {Astrophys. J.},
year = 2026,
volume = {1005},
number = {2},
pages = {182},
doi = {10.3847/1538-4357/ae7e87}
}

@ARTICLE{Kraichnan1965,
author = {{Kraichnan}, R.H.},
title = "{Inertial-Range Spectrum of Hydromagnetic Turbulence}",
journal = {Phys. Fluids},
year = 1965,
volume = 8,
pages = {1385-1387}}

@ARTICLE{Iroshnikov1964,
author = {{Iroshnikov}, P.S.},
title = "{Turbulence of a conducting fluid in a strong magnetic field}",
journal = {Soviet Astron.},
year = 1964,
volume = 7,
pages = {566-571}}

@ARTICLE{GaltierC2006,
author = {{Galtier}, S. and {Chandran}, B.~D.~G.},
title = "{Extended spectral scaling laws for shear-Alfv{\'e}n wave turbulence}",
journal = {Phys. Plasmas},
year = 2006,
volume = {13},
number = {11},
pages = {114505}}

@article{Costa2026,
  title = {Stability of Stationary Solutions in Acoustic Wave Turbulence},
  author = {Costa, Guillaume and Krstulovic, Giorgio and Nazarenko, Sergey},
  year = 2026,
  month = feb,
  journal = {Physical Review E},
  volume = {113},
  number = {2},
  pages = {024205},
  issn = {2470-0045, 2470-0053},
  doi = {10.1103/j1hs-b71h},
  urldate = {2026-05-05},
  langid = {english}
}

\end{document}